\documentclass[preprint,showpacs,preprintnumbers,amsmath,amssymb,nofootinbib]{revtex4}

\usepackage{etex}

\usepackage{amsthm,amscd,amsbsy,array}
\usepackage{bm}
\usepackage{soul} 

\usepackage{graphics,graphicx,xcolor}

\usepackage{soul} 

\usepackage[colorlinks=true, pdfstartview=FitV, linkcolor=blue, citecolor=blue, urlcolor=blue]{hyperref} 

\newcommand{\gb}{\colorbox{green}}

\newcommand{\dgreen}{\textcolor[rgb]{0,0.35,0}}

\newenvironment{redtext}{\color{red}}{\ignorespacesafterend} 
\newenvironment{bluetext}{\color{blue}}{\ignorespacesafterend} 
\newenvironment{greentext}{\color{green}}{\ignorespacesafterend}
\newenvironment{dgreentext}{\color{\dgreen}}{\ignorespacesafterend}
 
\newcommand{\bblue}{\begin{bluetext}} 
\newcommand{\eblue}{\end{bluetext}} 
\newcommand{\bred}{\begin{redtext}}
\newcommand{\ered}{\end{redtext}}
\newcommand{\bgreen}{\begin{greentext}}
\newcommand{\egreen}{\end{greentext}}
\newcommand{\bdgreen}{\begin{dgreentext}}
\newcommand{\edgreen}{\end{dgreentext}}
\numberwithin{equation}{section}

\let\ssection=\section
\renewcommand{\section}{\setcounter{equation}{0}\ssection}

\newcommand{\cA}{{\mathcal{A}_{+}}}

\newcommand{\tQ}{\widetilde{Q}}
\newcommand{\tL}{\widetilde{L}}

\newcommand{{\cL}}{{\mathcal{L}}}

\newcommand{\bp}{{\bf p}}

\newcommand{\bq}{{\bf q}}

\newcommand{\cQ}{\mathcal{Q}}

\def\smallover#1/#2{\hbox{$\textstyle\frac{#1}{#2}$}} %

\def\where{{\quad\text{where}\quad}}

\def\aand{{\quad\text{and}\quad}}

\def\bp{{\bm{p}}}

\def\bq{{\bm{q}}}
\def\parag{\hfil\break} 
\def\kikezd{\parag\underbar}
\def\benu{\begin{enumerate}}
\def\eenu{\end{enumerate}}
\def\beq{\begin{equation}}
\def\eeq{\end{equation}}
\def\beqa{\begin{eqnarray}}
\def\eeqa{\end{eqnarray}}

\def\barray{\left(\begin{array}}
\def\earray{\end{array}\right)}
\def\barraynb{\begin{array}}
\def\earraynb{\end{array}}
\def\ort{{\rm o}}

\def\?{\quad{\gb{\fbox{\texttt{?}}\;}}\quad}
\def\p{{\partial}}

\def\Rarrow{{\quad\Rightarrow\quad}}

\def\beq{\begin{equation}}
\def\eeq{\end{equation}}
\def\bea{\begin{eqnarray}}
\def\eea{\end{eqnarray}}

\def\p{\partial}

\def \p{{\partial}}

\def\G11{\Gamma_{11} }

\newcommand{\const}{\mathop{\rm const.}\nolimits}
\newcommand{\half }{\frac{1}{2}}

\def\smallover#1/#2{\hbox{$\textstyle\frac{#1}{#2}$}} %
\def\smallcirc{{\raise 0.5pt \hbox{$\scriptstyle\circ$}}}
\def\2{{\smallover1/2}}

\newcommand{\bigbox}[1]{\fbox{%
\rule[-20pt]{0pt}{45pt}$\;\;\displaystyle{#1}\;\;$}
}

\let\ssection=\section
\renewcommand{\section}
{\setcounter{equation}{0}\ssection}

\def\besub{\begin{subequations}}
\def\esub{\end{subequations}}

\begin{document} 

\preprint{\texttt{arXiv:1903.05070v5}}

\title{A generalized Noether theorem for scaling symmetry
\\[6pt]
}

\author{
P.-M. Zhang${}^{1,2}$\footnote{e-mail:zhangpm5@mail.sysu.edu.cn},
M. Elbistan${}^{2,3}$\footnote{mailto:mahmut.Elbistan@lmpt.univ-tours.fr},
P. A. Horvathy${}^{3}$\footnote{mailto:horvathy@lmpt.univ-tours.fr}
P. Kosi\'nski$^{4}$\footnote{email: pkosinsk@uni.lodz.pl}
}

\affiliation{
${}^{1}$ {\it School of Physics and Astronomy}, 
 \\ Sun Yat-sen University Zhuhai 519082, China
\\
${}^{2}$ {\it Institute of Modern Physics}
\\ Chinese Academy of Sciences,
 Lanzhou, China
\\
${}^3$ Institut Denis-Poisson CNRS/UMR 7013 - Universit\'e de Tours - Universit\'e d'Orl\'eans Parc de Grammont, 37200, Tours, (France)
\\
${}^4$
{\it Department of Computer Science}, Faculty of Physics and Applied Informatics\\
University of L\'od\'z,
Pomorska 149/153, 90-236 L\'od\'z, Poland
\\
}

\date{\today}


\begin{abstract} 
The recently discovered conserved quantity associated with Kepler rescaling is generalised by an extension of Noether's theorem which involves the classical action integral as an additional term. For a free particle the familiar Schroedinger-dilations are recovered. A general pattern arises for homogeneous potentials. The associated conserved quantity allows us to derive the virial theorem. The relation to the Bargmann framework is explained and illustrated by exact plane gravitational waves.  
\\[8pt]\noindent{}
{Eur. Phys. J. Plus (2020) 135:223\\
  doi.org/10.1140/epjp/s13360-020-00247-5 
}
\end{abstract}

\maketitle

\tableofcontents

\section{Introduction}\label{Intro}

A \emph{Noether symmetry} is a mapping $(\bq,t)\to (\bq',t')$  which leaves the Lagrangian $L$ invariant up to a surface term, 
\beq
{L}\big(\bq',\frac{d\bq'}{dt'},t'\big)\frac{dt'}{dt} \to {L}
\big(\bq,\frac{d\bq}{dt},t\big) + \frac{df(\bq,t)}{dt}\,.
\label{Oldsymmdef}
\eeq
Then the consequences are twofold:
\benu
\item
The classical Hamiltonian action $\displaystyle\int_0^t\!Ldt $ 
changes by a path-independent term, implying that the 
 variational equations remain invariant and thus a symmetry carries a motion into another motion.

\item
Noether's theorem associates a conserved quantity  to each continuous group of symmetries.

\eenu

A classical ``counter-example''  when one, but not both consequences hold is provided by the  rescaling of position and time in Kepler the problem~:
\besub
\begin{align}
&t \to \lambda^3\, t\,,\qquad\qquad
\bq \to \lambda^2\,\bq\,, \qquad\qquad
\lambda=\const
\label{K3finite}
\\
&\delta t \to 3(\delta\lambda)\, t\,,\qquad
\delta \bq \to 2(\delta\lambda) \,\bq\,, \qquad\;
\delta\lambda=\const\,,
\label{K3infin}
\end{align}
\label{K3}
\esub
which takes  planetary trajectories  into planetary trajectories, whereas the  Lagrangian changes by a constant factor,
\beq
\label{KLag} 
{L}_{Kepler}=\frac{m}{2}\left(\frac{d{\bq}}{dt}\right)^2+\frac{GmM_{\odot}}{|\bq|}\,
\quad \to \quad
\lambda^{-2}\, {L}_{Kepler}\,.
\eeq
 The rescaling (\ref{K3}) is therefore  \emph{not a symmetry} for the Keplerian system in the above sense and therefore no Noetherian conserved quantity is expected to arise ; textbooks call it a ``similarity" \cite{LLMech}.   
It came therefore as a surprise  that the Kepler problem \emph{does} have a conserved quantity associated with (\ref{K3}) -- which is however of a nonconventional form, involving also the classical Hamiltonian action \cite{KHarmonies},  
\beqa 
Q_{Kepler} =
 m \frac{\;\, d}{dt}(\bq^2) - 3 t E - S(t)\,,
 \qquad
S(t)=\int_0^t \!\!\!{L}_{Kepler} \,d\tau\, ,
\label{QKepler}
\eeqa
where the integration is along the classical trajectory in $3$-space. 

This novel conserved quantity which seems to have escaped attention until recently was obtained  in \cite{KHarmonies} in a remarkably indirect way~: the Kepler problem was first ``E-D'' (Eisenhart-Duval) lifted to a 5-dimensional ``Bargmann'' manifold 
\cite{Eisenhart,Bargmann,DGH91}. See sec.\ref{GWSec} for details.

The aim of this paper is to generalise this type of quantity by \emph{extending Noether's theorem} first within the framework of analytical mechanics. 
Applications include, besides planetary motion, homogeneous potentials. 

The relation to the Bargmann framework is explained in sec. \ref{GWSec}. Particular attention is devoted to exact plane gravitational waves which correspond to time-dependent anisotropic oscillators \cite{Conf4GW}.
  
\section{A generalized Noether theorem}

Let us assume that we have a dynamical system with generalized coordinates $q_i, \, i=1,\dots, n$ and a Lagrangian $L = L(\bq,\dot{\bq},t)$ and consider a $1$-parameter family of transformations  $\bq\to\bq'(t')\,,t\to t'$, obeying
\beq\bigbox{
{L}\big(\bq',\frac{d\bq'}{dt'},t'\big)\frac{dt'}{dt} = \Lambda\,{L}
\big(\bq,\frac{d\bq}{dt},t\big) + \frac{df(\bq,t)}{dt}
}
\label{Gensymmdef}
\eeq
where $\Lambda$ is some constant and $f$ an arbitrary (differentiable) function. Then the associated Lagrange equations are invariant. Anticipating  the terminology of Duval et al \cite{DThese} discussed in sec. \ref{GWSec} such transformations 
will be called \emph{Chrono-Projective}.

Then, proceeding in the standard way allows us to deduce from (\ref{Gensymmdef}) the identity
\beq
\frac{\,d }{dt}\bigg(
\frac{\p{L}}{\p\dot{q}_i}\delta{q}_i-
\big(\frac{\p{L}}{\p\dot{q}_i}\dot{q}_i-L\big)\delta{t}-\delta{f}\bigg)
+\Big(\frac{\p{L}}{\p{q}_i}-\frac{\,d }{dt}\big(\frac{\p{L}}{\p\dot{q}_i}\big)\Big)
\big(\delta{q}_i-\dot{q}_i\delta{t}\big)-
(\delta{\Lambda})\,L=0\,.
\label{LongId}
\eeq

This relation is converted into a conservation law as follows. One solves the Lagrange equations of motion on some interval
$0\leq\tau\leq t$ with the ``initial'' [in fact final] conditions $q_i(t)=q_i$ and $\dot{q}_i(t)=\dot{q}_i$. Then the Hamiltonian action calculated along the trajectory,
\beq
S=\int_0^t\!\!L(\bq(\tau),\dot{\bq}(\tau),\tau)\,d\tau\,,
\label{Saction}
\eeq
becomes a function of the end points $q_i,\,\dot{q}_i$ and $t$ ; moreover, ${dS}/{dt}= L\,.$ 
Inserting the Hamiltonian  and using the  Lagrange equations, eqn. (\ref{LongId}) allows us deduce the conserved charge,
\beq
\bigbox{
Q=\frac{\p{L}}{\p\dot{q}_i}\delta{q}_i-
H\delta{t}-\delta{f}-(\delta{\Lambda})\,\int_0^t\!\!Ld\tau\,.
}
\label{newconsQ}
\eeq
Putting here (\ref{K3infin}) and $\delta{f}=0$ yields, for example, the Kepler charge (\ref{QKepler}).

More generally, let's assume that (\ref{Gensymmdef}) is satisfied (with $f=0$ for simplicity) and consider the Kepler-type rescaling 
\besub
\begin{align}
&q_i' = \lambda^a q_i \approx q_i+ a\,q_i\,\delta\lambda\,,
\label{qscale}
\\
&t'=\lambda^bt \;\; \approx t+ b\,t\,\delta\lambda\,,
\label{tscale}
\\
&\Lambda= \lambda^c \;\,\approx 1+c\,\delta\lambda\,
\label{Lambdascale}.
\end{align}
\label{abcscale}
\esub
where $a, b, c$ are to be determined. ($\approx$ means to first order).

Let us assume that the Lagrangian is of the form
$ 
L=g_{ij}\dot{q}_i\dot{q}_j-V(\bq),
$ 
where for simplicity we took $g_{ij}$  a symmetric constant matrix. This Lagrangian(times $dt$) scales under (\ref{qscale})-(\ref{tscale}) as
\beq
L\,dt \to \lambda^{2a-b}\Big[g_{ij}\dot{q}_i\dot{q}_j-\lambda^{-2a+2b}V(\lambda^a\bq)\Big]dt\,.
\label{Ldtscale}
\eeq
Let us first turn off the potential, $V(\bq)\equiv0$, i.e., 
$
L=L_{free}=\half\dot{\bq}^2\,.
$
Then (\ref{Ldtscale}) says  that 
$ 
L_{free} \to \lambda^{2a-b}L_{free} \,.
$
The generalized symmetry condition (\ref{Gensymmdef}) is thus satisfied with 
$
\Lambda = \lambda^c \,\text{i.e. if}\, c=2a-b\,.
$
For any choice of $a$ and $b$
  (\ref{newconsQ}) associates a conserved charge, namely
\beq
Q =  a\, 2 g_{ij}\dot{q}^i q^j
-btH-c\int_0^t\!\!Ld\tau\,,
\qquad
c=2a-b.
\label{abQ}
\eeq 
But $H_{free}=L_{free}$ and therefore $b$ drops out, and the last two terms combine into a single one~:
\beq
Q_{free} = a(\dot{\bq}\cdot\bq-\dot{\bq}^2t)=
a\dot{\bq}\cdot\big(\bq-\dot{\bq}t\big)\,,
\label{Schrdil}
\eeq
where we recognise the scalar product of the (separately conserved) \emph{momentum} with the \emph{center of mass}. 
(\ref{Schrdil}) quantity is actually the conserved quantity associated with \emph{Schr\"odinger dilations} \cite{JacobiLec,Schr,DGH91,Conf4GW},
$
Q_{free} = D=(\dot{\bq}\cdot\bq-2tH_{free})\,,
$
traditionally obtained when time is dilated twice as much as space, i.e., $b=2a$.

 In the free case, this is the end of the story.
Let us now restore the potential, $V$.  The overall scaling $\Lambda$ is already determined by the kinetic term then (\ref{Gensymmdef}) requires, 
\beq
\lambda^{-2(a-b)}V(\lambda^a\,\bq)=V(\bq)\,.
\eeq
For homogeneous potentials,  $V(\mu\bq)=\mu^{k}V(\bq)$,  the symmetry condition requires
\beq
(-2+k)a + 2b =0
\Rarrow
c=a\big(1+\frac{k}{2}\big)\,.
\label{ksymm}
\eeq
The integral term contributes whenever $k\neq-2$ ; the value of (\ref{abQ}) is
$ 
Q=2a\,g_{ij} q^i(0) \dot{q}^j(0)\,.
$ 
  
When $a\neq0$ $z=b/a$ is the  \emph{dynamical exponent} \cite{Henkel,DHNC}) ; then $a=1$ can be chosen and the integral in (\ref{abQ}) has coefficient $c=2-z$. 

\benu
\item
For \emph{free motion} $k=0$, $a=b=c$, the terms combine, and we get (\ref{Schrdil}).

\item
For the \emph{inverse-square potential} $V\propto |\bq|^{-2}$ eqn.  (\ref{ksymm}) requires
$c=0$ i.e. $z=2$ and the Schr\"odinger scaling \cite{Schr,Fubini,JacobiLec} is recovered.

\item
For the \emph{Newtonian potential} $V\propto |\bq|^{-1}$ we have $k=-1$  and then get the Kepler scaling, 
 $z=3/2$, cf. (\ref{K3infin}) ; the  integral term does \emph{not} drop out from the Kepler charge (\ref{QKepler}) \cite{KHarmonies}.

\item
For the \emph{constant force}  $k=1$ ; the scaling is
\beq
t\to \lambda \, t, \qquad \bq \to \lambda^2 \bq,
\qquad
S \to \lambda^3S\,.
\label{linforcescale}
\eeq
It is amusing to figure that we climb, with Galilei, to the top of the Pisa Leaning Tower and drop a stone from $q(0)=0$.
The conserved charge is
\beq
Q_{Gal}=2\, 
\bq(t)\cdot\dot{\bq}(t)
-3\int_0^t\! L_{Gal}d\tau\,,
\quad
L_{Gal}=\half\dot{q}^2-q
\Rarrow
Q_{Gal}=0\,.
\label{Galcharge}
\eeq

\item
For a \emph{harmonic oscillator} $k = 2$ ;  (\ref{ksymm}) implies  $b=0$, i.e., pure position-rescaling. 
\beq
Q_{osc} = \bq(t)\cdot\dot{\bq}(t) - 2 \int_0^t L_{osc} d\tau,
\quad
L_{osc}=\half \dot{\bq}^2-\half \omega^2\bq^2
\Rarrow Q_{osc} = q_i(0) \dot{q}_i(0)
\,.
\label{osciQ}
\eeq
\eenu

Let us study the oscillator case in some detail.
 After a full period $t=T=2\pi/\omega$ we are back where we started from, so  the first term in (\ref{osciQ}) vanishes. Therefore
\beq
\frac{1}{T}
\int_0^{T}\half\dot{\bq}^2dt = \frac{1}{T}\int_0^{T} \half \omega^2\bq^2dt\,,
\label{virial}
\eeq
i.e., the average over a period of the kinetic energy equals to the 
average of the potential energy -- which is the \emph{virial theorem} for an oscillator. 

The virial theorem can actually be generalized to any $k$ along the same lines. For the standard Lagrangian/Hamiltonian $L=K-V$ resp. $H =K+V$
the associated conserved quantity (\ref{abQ}) can be written as 
\beq
Q =  a\,q_i(t)\,\dot{q}_i(t)
-2a\int_0^t\!\!Kd\tau\, +2(a-b)\int_0^t\!\!Vd\tau\,,
\label{abQKV}
\eeq
whose conservation implies for a periodic motion with period $T$ 
\beq
< K > \;=\;\frac{1}{T}\int_0^T K d\tau = 
\dfrac {k}{2}\,\frac{1}{T} \int_0^T V d\tau
\;=\;
 \dfrac{k}{2} <V>\,.
\label{genvir}
\eeq
We conclude this section by presenting an alternative point of view related to \emph{shape invariance}. Let us assume that our Lagrangian $\tL$  depends on an additional parameter  $\mu$. Let a transformation $\bq'=\bq'(\bq,t)$ and $t'=t'(\bq,t)$ be completed by $\mu'=\mu'(\mu)$ and assume that we have, instead of (\ref{Gensymmdef}),
\beq
{\tL}\big(\bq',\frac{d\bq'}{dt'},t',\mu'\big)\frac{dt'}{dt} = {\tL}
\big(\bq,\frac{d\bq}{dt},t,\mu\big) + \frac{df(\bq,t)}{dt}\,.
\label{Gensymmdefbis}
\eeq
Proceeding as before, we find the modified identity 
\beq
\frac{\,d }{dt}\bigg(
\frac{\p{\tL}}{\p\dot{q}_i}\delta{q}_i-
\big(\frac{\p{\tL}}{\p\dot{q}_i}\dot{q}_i-\tL\big)\delta{t}-\delta{f}\bigg)
+\Big(\frac{\p{\tL}}{\p{q}_i}-\frac{\,d }{dt}\big(\frac{\p{\tL}}{\p\dot{q}_i}\big)\Big)
\big(\delta{q}_i-\dot{q}_i\delta{t}\big)
+\frac{\p{\tL}}{\p\mu}\delta\mu
=0\,,
\label{LongIdbis}
\eeq
which yields a ``partial conservation law'' for the \emph{usual} conserved charge $\tQ$,
\beq
\tQ= \frac{\p{\tL}}{\p\dot{q}_i}\delta{q}_i-
H\delta{t}-\delta{f},
\qquad
\frac{d\tQ}{dt}+\frac{\p{\tL}}{\p\mu}\delta\mu=0
\,.
\label{parCL}
\eeq
Choose in particular $\tL=\mu\,L$, where $L$ is assumed to satisfy  (\ref{Gensymmdef}) with $\Lambda=1$ under the rescaling (\ref{abcscale}). The Lagrangians $L$ and $\tL$ yield, for each fixed value of $\mu$, identical equations of motion. Our clue is that supplementing (\ref{abcscale}) with
\beq
\mu'=\lambda^{-c}\mu\,,
\label{muscale}
\eeq
the modified Lagrangian $\tL$ becomes invariant. Moreover, 
$(\p{\tL}/\p\mu)\delta\mu=c\mu\tL\delta\lambda=\frac{\;d}{dt}\!\displaystyle\int^t\!\tL d\tau$ and therefore, putting ${\mu=1}$,  (\ref{parCL}) allows us to recover
the modified conserved quantity (\ref{newconsQ}) as  $Q=\tQ-(\p{\tL}/\p\mu)$.


\section{Hamiltonian framework}

In the Lagrangian framework, the Noether theorem applies to point transformations only.
However it can also be reformulated in the Hamiltonian framework, leading to conserved charges generated by canonical symmetry transformations. 
The usual relation defining canonical transformations can be generalized in fact to include the scale factor 
$\Lambda$ as follows. 
The action integral becomes, after a Legendre transformation,
\beq
\int\big[p_i'\dot{q}_i'-H'(\bq',\bp',t)\big]dt =
\int \Big[\Lambda\,\big(p_i\dot{q}_i-H(\bq,\bp,t)\big)-\frac{d\Phi}{dt}\Big]\,dt\,.
\label{HamHamact}
\eeq
Putting $\Psi(\bq,\bp')=\Phi+q_i'p_i'$, this can be rewritten as
\beq
\int\big[-q_i'\dot{p}_i'-H'\big]dt =
\int\Big[\Lambda\,(p_i\dot{q}_i-H)-\frac{d\Psi}{dt}\Big]\,dt\,,
\label{HamHamact2}
\eeq
which yields \vspace{-4mm}
\besub
\begin{align}
&p_i=\frac{1}{\Lambda}\frac{\p\Psi(\bq,\bp',t)}{\p{q}_i}\,,
\\
&q_i'=\frac{\p\Psi(\bq,\bp',t)}{\p{p}_i'}\,,
\\
&H'=\Lambda H+\frac{\p\Psi(\bq,\bp',t)}{\p{t}}\,.
\end{align}
\label{cantransf}
\esub
The identity transformation is generated by the function
$\Psi_0=q_ip_i'$ and $\Lambda=1$. Therefore the infinitesimal transformation is obtained by putting 
$\Lambda = 1 +\delta{\Lambda}$ and
\beq
\Psi(\bq,\bp',t)=\Psi_0(\bq,\bp')+\delta G(\bq,\bp',t)
=\Psi_0(\bq,\bp')+\delta {G}(\bq,\bp,t)
\label{PsiG}
\eeq
where in  we replaced $\bp'$ by $\bp$ because $\delta{G}$ is already infinitesimal. 
Then our eqns yield
\besub\begin{align}
&\delta\!p_i=p_i'-p_i=\delta{\Lambda}\,p_i
-\frac{\p\,(\delta{G})}{\p{q}_i}\,,
\\
&\delta\!q_i=q_i'-q_i=\frac{\p\,(\delta{G})}{\p{p}_i}\,,
\\
&H'=H+\delta{\Lambda}\,H+\frac{\p\,(\delta{G})}{\p{t}}\,.\qquad
\end{align}
\label{deltapqH'}
\esub
A canonical transformation is a symmetry if the 
Hamiltonian equations are form-invariant,
$ 
H'(\bq',\bp',t)= H(\bq',\bp',t)\,.
$ 
Expanding  to the first order and using
(\ref{deltapqH'}) one finds,
\beq
\Big\{\delta{G},H\Big\}+\frac{\p(\delta{G})}{\p{t}} 
+\delta{\Lambda}\,\big(H-p_i\frac{\p H}{\p{p}_i}\big)=0
\;\;\text{i.e.,}\;\;
\bigbox{
\frac{\, d}{d{t}}\left(\delta {G} - \delta{\Lambda}\,\int_0^t\!\!Ld\tau\right)=0
}
\label{genHacharge}
\eeq
which is the generalized Noether charge.

For a point transformation we get, for $\delta f=0$, 
\beq
\delta\,G = \frac{\p L}{\p \dot{q}_i}\delta{q}_i-H\delta{t}=p_i\delta{q}_i-H\delta{t}\,.
\eeq
In the Lagrangian framework time and space transformations can be combined. However time is fixed in the Hamiltonian one, and so in order to include time-variable transformations, we have to replace $\delta{q}_i, \, \delta{p}_i $ by
\beq
\barraynb{lllll}
\delta_H{q}_k&=&\big\{q_k,\delta{G}\big\}
&=&
\delta{q}_k-\dot{q}_k\delta{t}
\\[6pt]
\delta_H{p}_k&=&\big\{p_k,\delta{G}\big\}
&=&
-p_i\displaystyle\frac{\p(\delta{q}_i)}{\p{q}_k}-\dot{p}_k\delta{t}
\earraynb
\eeq
i.e., for point transformations, $q_k$ and $p_k$ are corrected by a time shift.
But in the Hamiltonian framework Noether's theorem is more general~:  $\delta{G}$  can have a more complicated form, and generate canonical transformations which can not be derived from point transformations.

We note that for the scaling transformations (\ref{abcscale}) the generator
$\delta{G}$ in (\ref{PsiG})-(\ref{deltapqH'})  could be expressed in terms of
the initial conditions as $\delta{G}=a q_i(0)\dot{q}_i(0)\delta\lambda$.

\section{Chrono-projective symmetries and gravitational waves}\label{GWSec}

A convenient way to study non-relativistic conformal symmetries is to use the ``Bargmann'' framework  \cite{Bargmann,DGH91}~: one lifts the non-relativistic dynamics in $(d,1)$ dimensions to a $(d+2)$ dimensional manifold $M$ endowed with a Lorentz metric $g_{\mu\nu}$ and a covariantly constant null vector $\xi$, referred to as a Bargmann space \cite{Bargmann,DGH91}. The original non-relativistic motions  are projections of null geodesics lying ``upstairs'', i.e., in Bargmann space. For motion in a potential $V(q,t)$ the appropriate metric resp. ``vertical vector'' are 
\beq
g_{\mu\nu}dx^\mu dx^\nu=d\bq^2 +2dtds-2Vdt^2
\aand
\xi = \p_s\,.
\label{Bmetricxi}
\eeq
The vertical vector $\xi$ generates and isometry, whose conserved momentum associated with the Killing vector $\p_s$ upstairs,  $P_s$ is the physical mass ``downstairs'' i.e., in $2D$ non-relativistic spacetime.

The remarkable fact recognized already by Eisenhart \cite{Eisenhart} is that the null lift  of a non-relativistic motion (we call the Eisenhart-Duval (E-D) lift) is, 
\beq
\big(q^i(t),t, s_0-S(t)\big)
\where
S(t)=\int_0^t\! L_{cl}(q(\tau),\tau,)d\tau
\label{Elift}
\eeq
i.e., the ``vertical'' coordinate, $s$ is essentially i.e. up to a constant \emph{minus
the classical Hamiltonian action} calculated along the trajectory. It is precisely this rule (\ref{Elift}) which guarantees that the lift is null, as shown by evaluating (\ref{Bmetricxi}) for a tangent vector.

Conformal transformations of $(M,g_{\mu\nu})$ take  null-geodesics into  null-geodesics; however such a  transformation generated by a vector field $Y$  projects to a symmetry for the non-relativistic dynamics ``downstairs" only if it satisfies the additional condition
\beq
L_Y\xi = 0.
\label{Schrcond}
\eeq
For the Minkowski metric written in light-cone coordinates, for example, (\ref{Bmetricxi}) with $V=0$,
 the conformal algebra is $\ort(d+1,2)$ and those vectorfields which satisfy (\ref{Schrcond}) span the centrally extended Schr\"odinger group \cite{Bargmann,DGH91}. Schr\"odinger dilations are, in particular, identical to $Q_{free}$ in (\ref{Schrdil}).

The Kepler rescaling (\ref{K3}) can  be lifted to Bargmann space [with $V=-GM_{\odot}/|\bq|$],  as
\beq
t\to \lambda^3t,\; \bq \to \lambda^2,
\;
s\to \lambda s
\qquad\text{\small infinitesimally}\qquad
Y=3t\p_t+2\bq\cdot\p_{\bq}+s\p_s\,,
\label{BKeplerscale}
\eeq
 becoming there  a  \emph{conformal transformation}; the scaling of the coordinate $s$ is dictated precisely by this requirement \cite{KHarmonies,Conf4GW,Cariglia15}.  
 
 However  (\ref{BKeplerscale}) does \emph{not} satisfy (\ref{Schrcond})~: $L_Y\xi=-\delta\lambda\, \xi$. 
Transformations of the Bargmann space such that $L_Y\xi$ is \emph{parallel} to $\xi$,
\beq
L_Y\xi = \psi\, \xi
\label{ChronoCond}
\eeq
for some (necessarily constant) $\psi$ are in fact the lifts of \emph{chrono-projective transformations}, originally introduced in terms of the Newton-Cartan structure ``downstairs \cite{DThese,5Chrono}. These same condition was also considered, independently, by Hall et al \cite{HallSteele} in their study of conformal transformations for gravitational waves.

Conformal transformations are symmetries for null geodesics \emph{upstairs} and generate there, through   Noether's theorem, conserved quantities
 -- but the associated conserved charge may \emph{not} project to ordinary space-time as well-defined quantities. However, when the additional ``Chrono-projective'' condition (\ref{ChronoCond}) is satisfied, then they fail  to project by just a little: 
 it is enough to subtract a constant term proportional to the initial value $s_0$ to get a perfectly well-defined conserved quantity for the projected dynamics \cite{KHarmonies,Conf4GW}. 
This is  what happens for (\ref{K3})~: eqn. (\ref{Elift}) implies that
\beq
Q_{Kepler}=\cQ_{Kepler}-ms_0
\where
\cQ_{Kepler}=-3tE+2m\bq\cdot\dot{\bq}+ms
\label{QKeplerbis}
\eeq
consistently with (\ref{QKepler}).

\kikezd{1-dim oscillator}.
 An even simpler illustration is provided by a 1-d harmonic oscillator. Its  3d Bargmann space has the metric is $ds^2=dq^2-2dtds-\omega^2q^2dt^2$ and $\xi=\p_s$. 
\begin{figure}
\includegraphics[scale=.15]{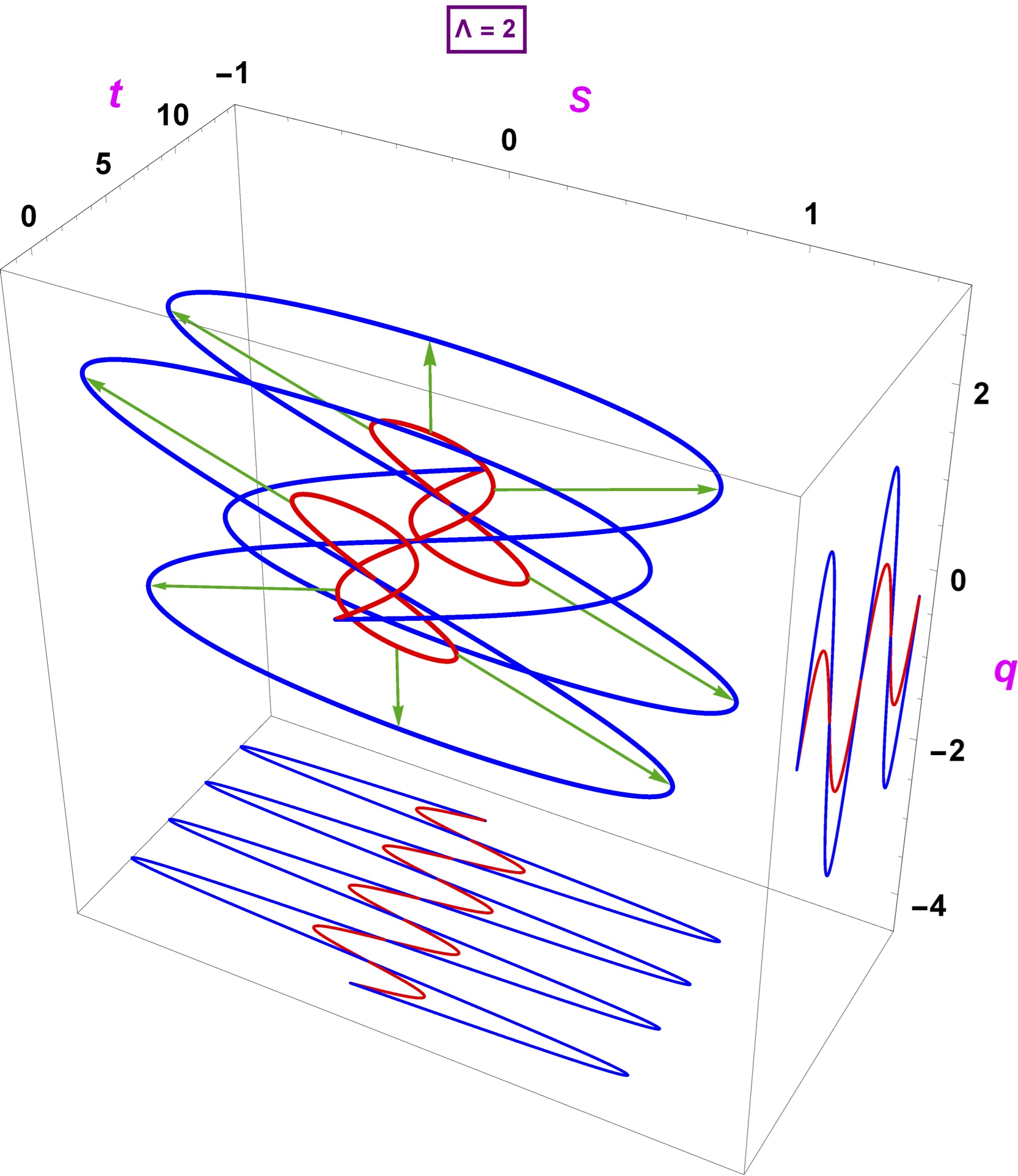}\vskip-4mm
\caption{\textit{\small 
The  Eisenhart-Duval lifts to $3d$ Bargmann space of two motions of a 1d harmonic oscillator. Projected to  the $(q,t)$ plane [which is here vertical] we get the familiar sinus curves related by position-alone scaling by $\lambda$. The Hamiltonian actions calculated along the trajectories, shown in the $(S \, , \, t)$ [here horizontal] plane oscillate with double frequency and are scaled by $\lambda^2$. The E-D lift of the scaling [in \dgreen{\bf green}], (\ref{osciscalelift}), the homothety 
carries the lifted curves into each other.
}}
\vskip-1mm
\label{osciBfig}
\end{figure}
The space-alone rescaling  lifts to $3d$ Bargmann space as the \emph{homothety},
\beq
t\to t, \qquad {q}\to \lambda\, {q},\qquad s \to \lambda^2 s
\label{osciscalelift}
\eeq
and generates there a chrono-projective symmetry for null geodesics \cite{5Chrono,DGH91,KHarmonies,Conf4GW}, as shown on Fig.\ref{osciBfig}.
 The associated conserved quantity is (\ref{osciQ}).

 Let us consider two motions ${q}_1(t)$ and ${q}_2(t)$  which start from the same initial position $q_1(0)=0=q_2(0)$ but with different initial velocities,
$
\dot{q}_2(0)=\lambda \dot{q}_1(0).
$
The space-alone rescaling
$ 
q\to \lambda\, q,\, t\to t
$ 
takes ${q}_1(t)$ into ${q}_2(t)$.
Both motions return to their initial position after a half period (since the latter is independent of the initial conditions, as purportedly observed by Galilei in the Pisa cathedral).

\section{Gravitational waves and oscillators}

\begin{figure}[h]
\includegraphics[scale=.18]{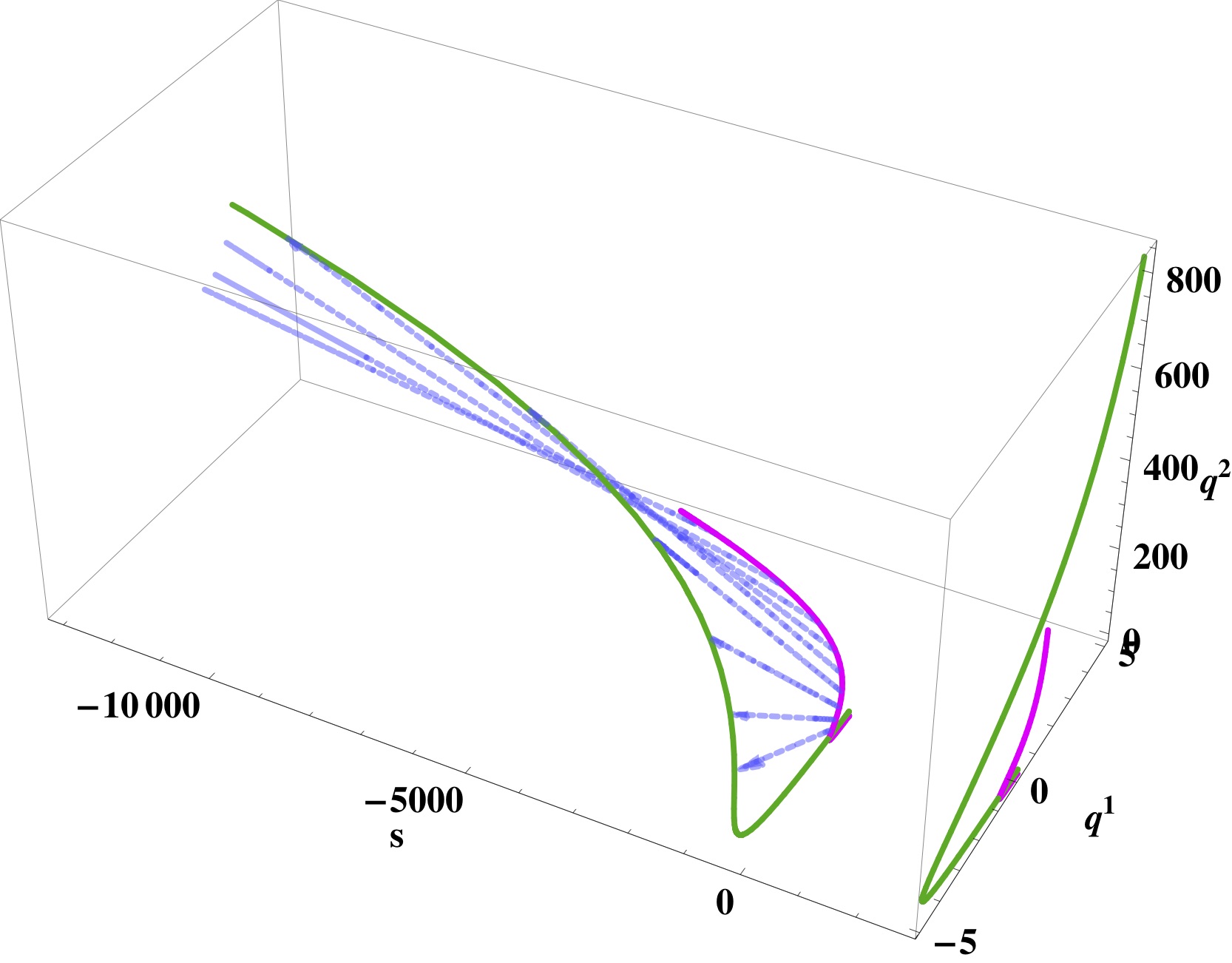}
\vskip-3mm
\caption{\textit{\small The Brdi\v{c}ka metric (\ref{Brdmetric}) provides us with the Bargmann description of a particle moving in a saddle  potential. 
The trajectory combines oscillation in the attractive $q^1$ sector with exponentially escaping motion in the repulsive $q^2$ and $s$ sectors.
The trajectories are carried into each other by the homothety (\ref{homothety}).
}}  
\label{Brdifig}
\end{figure}

Let us now consider the exact gravitational plane wave studied by Brinkmann \cite{Brinkmann}, 
\begin{subequations}
\begin{align}
g_{\mu\nu}dx^\mu dx^\nu=d\bq^2  
 + 2 dt ds + K_{ij}(t) q^i q^j dt^2\,,
\label{Bmetric}
\\[3pt]
K_{ij}(t){q^i}{q^j}=
\half{\cA}(t)\Big((q^1)^2-(q^2)^2\Big)+{\mathcal A}_{\times}(t)\,q^1q^2\,,
\label{Bprofile}
\end{align}
\label{genBrink}
\end{subequations}
where $\cA$ and ${\mathcal A}_{\times}$ are the $+$ and $\times$ polarization-state amplitudes. A simple example 
 is given by the $t$-independent linearly polarized gravitational wave \cite{Brdicka} with $\cA=2\Omega^2,\, \Omega = \const$ and ${\mathcal A}_{\times}=0$, i.e.,
\beq
g_{\mu\nu}dx^{\mu}dx^{\nu}
=d\bq^2 +2dtds-\Omega^2\Big(({q^1})^2-({q^2})^2\Big)dt^2\,.
\label{Brdmetric}
\eeq
Viewed as a Bargmann space, this metric describes an attractive oscillator in the $q^1$ coordinate combined with a repulsive (inverted) one in the $q^2$ sector ; $t$ corresponds to non-relativistic time (as anticipated). 
 The motion is governed by the geodesic Lagrangian
\beq
\half\left(({\dot{q}^1})^2+({\dot{q}^2})^2\right)+\dot{t}\dot{V}-\half\Omega^2\Big((q^1)^2-(q^2)^2\Big)\,
\label{BrdLag}
\eeq
where the dot denotes derivation w.r.t. an affine parameter.
For a particle initially at rest (e.g. with initial conditions
$q^1(0)= q^2(0)=1\; \dot{q}^1(0)= \dot{q}^2(0)=0$)
the  trajectory is, 
\beq\left\{\barraynb{lll}
q^1(t)=\cos\Omega t
\\[4pt]
q^2(t)=\cosh\Omega t
\earraynb\right.
\qquad
s(t)=s_0+\frac{\Omega}{4}\Big(
\sin 2\Omega t-\sinh 2\Omega t\Big)\,,
\label{Brdtraj}
\eeq
shown on fig.\ref{Brdifig}. 
Then the homothety (\ref{osciscalelift}) \cite{Torre,AndrPrenc,Conf4GW} 
generated by
\beq
Y_{hom}= q^i\p_{i}+2 s\p_{s}\,,
\label{homothety}
\eeq
is a \textit{chrono-projective transformation} \cite{5Chrono} of the metric (\ref{Brdmetric}),
\beq
L_{Y_{hom}}g=2\chi{g},\qquad
L_{Y_{hom}}\xi=-2\chi\xi\,.
\eeq
The associated  conserved  charge is
\besub
\begin{align}
&Q_{hom}=\cQ-2s_0=
\bq(t)\cdot\dot{\bq}(t)
-2S(t),
\label{OsciChronocharge}
\\[6pt]
&S(t)=\!\int_0^t\!\! L_{osc}(\bq(\tau), \dot{\bq}(\tau)) d\tau\,,
\qquad
L_{osc}=\half \dot{\bq}^2-\half \Omega^2\big((q^1)^1-(q^2)^2\big)\,.
\label{OsciAction}
\end{align}
\label{OsciChrono}
\esub
Evaluating the integral,
$Q_{hom}=(\bq\cdot\dot{\bq})(0)=0$.
\goodbreak

\begin{figure}[h]
\hskip-12mm
\includegraphics[scale=.21]{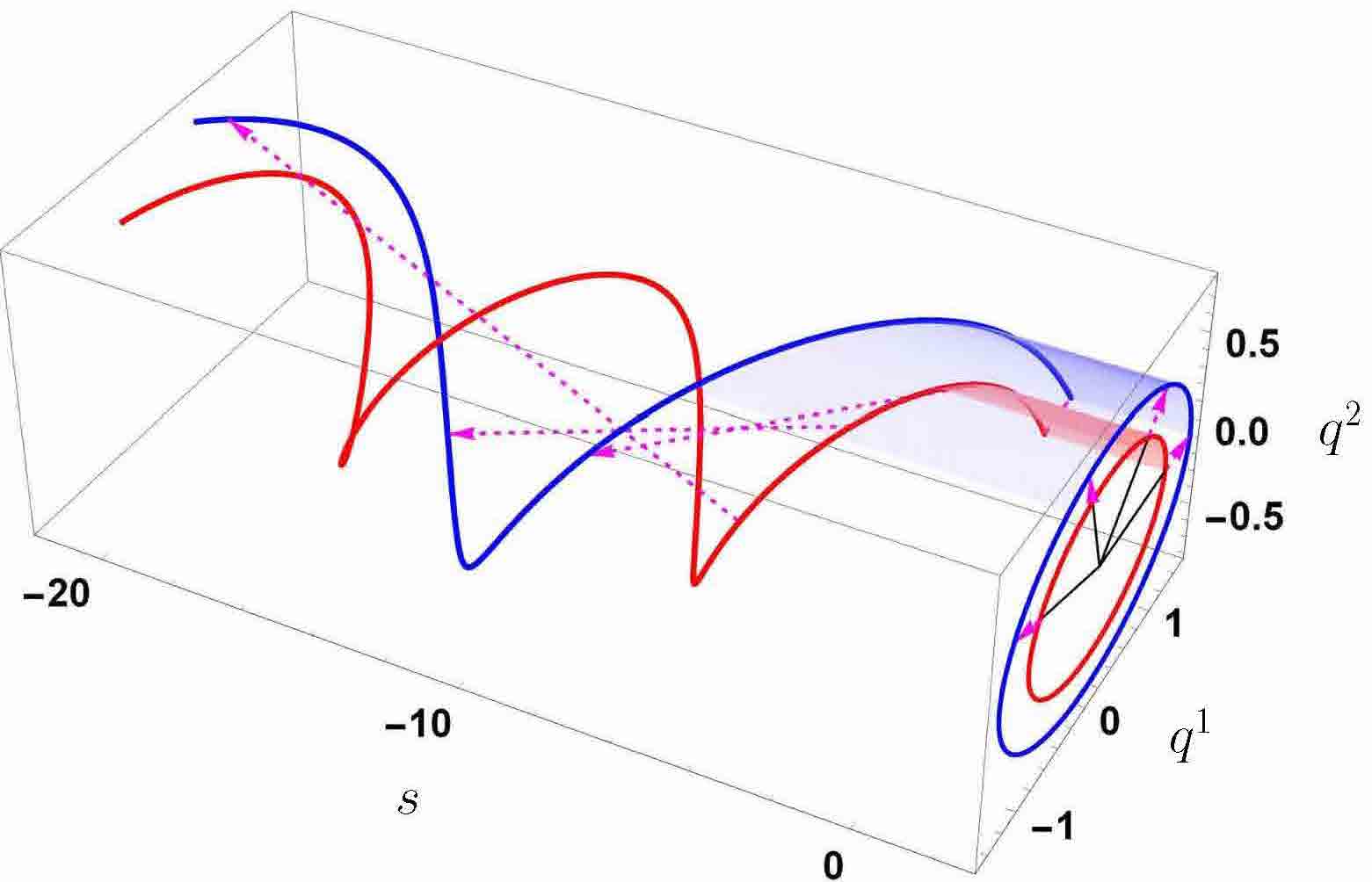}
\\
\null\vskip-12mm
\caption{\textit{\small  Motion of a $2d$ oscillator  unfolded to 4D Bargmann space [and dropping the non-relativistic time coordinate]. The homothety (\ref{homothety}) carries lifted oscillator-ellipses to lifted oscillator-ellipses.  
\label{OsciBarg}
}}
\end{figure}

We remark that changing the relative sign from minus to plus in (\ref{Brdmetric}), 
\beq
K_{ij}(t){q^i}{q^j}=
\half\Omega^2\Big((q^1)^2+(q^2)^2\Big),
\qquad
\Omega=\const
\label{isoosci}
\eeq
 would yield instead the Bargmann description of a \emph{time independent isotropic  harmonic oscillator} \footnote{ (\ref{isoosci}) is only a  pp wave, not a vacuum solution of the Einstein equations \cite{Brinkmann,DGH91}.}.
The familiar elliptic trajectories in the transverse plane lift to $4D$ Bargmann space as  null geodesics ; the $\xi$-preserving isometries span the centrally extended Newton-Hooke group, whose $\xi$-preserving conformal extension spans a group isomorphic to the Schr\"odinger group, etc. Here we just mention  that the  \emph{homothety} (\ref{homothety}) 
acts as a chrono-projective symmetry. The associated Noether charge is still $\cQ_{hom} = q^i P_i+2s P_s$, yielding the projected conserved quantity (\ref{OsciChronocharge})
\beq
Q_{osc} = q^iP_i-2\!\int_0^t\!\!L_{osc} d\tau
=\bq(0)\cdot \dot{\bq}(0)
\eeq 
  with  the isotropic oscillator Lagrangian $L_{osc}=\half \dot{\bq}^2-\half \Omega^2\bq^2$. 
  Fig.\ref{OsciBarg} should be compared with the Kepler figure  in \cite{KHarmonies}.

\section{Discussion}

In this paper we extended Noether's theorem to more general symmetry transformations which include also rescalings. Our results confirm the conserved charge found recently \cite{KHarmonies} for the Kepler problem; for a free particle we recover Schr\"odinger dilations \cite{JacobiLec,Schr}. For homogeneous potential (which include free fall and a harmonic oscillator) we get a new charge, whose conservation allows us we rederive the virial theorem. Further applications (namely to gravitational plane waves)  will be presented elsewhere \cite{Conf4GW}.

Compared with the usual Noether theorem, our new charge (\ref{newconsQ}) has has an extra term, in fact the classical action, which can be calculated only after the equations of motion had been solved. The new conservation law has nevertheless useful applications, as demonstrated by its use to prove the virial theorem and to derive Kepler's Third law \cite{KHarmonies}.

Further applications and generalizations are discussed in \cite{Conf4GW} where it is argued that a similar  conserved quantity arises for exact plane waves, and behaves as a symmetry for \emph{null} geodesics motion
\cite{Carroll4GW,POLPER,Ilderton}. The extension to massive geodesics is considered in \cite{Dimakis}.

We just mention that  Maxwell's electromagnetic Lagrangian under duality transformations would provide a field-theoretical example with helicity as associated conserved quantity.

After our paper was first posted to arXiv, our attention was called to  earlier investigations  \cite{vKampen, Nachtergaele}.
The closest  to our approach is that van Kampen  \cite{vKampen}, who uses also a Lagrangian framework ; his equation \# (5) is in fact our eqn. (\ref{Gensymmdef}). He could have [but did not] obtain our new charge (\ref{newconsQ}) by integrating his unnumbered equation after his \# (5) on p.237.

Nachtergaele et al. \cite{Nachtergaele} studied canonical transformations in the Hamiltonian framework in symplectic space, and applied them to Toda chains. 

Both of these papers focus on the virial theorem related to the Kepler rescaling ; no additional conserved quantity was found, though. 

We also came across \cite{Keplar} which uses Lie transformations and would also allow to derive our new charge (\ref{newconsQ}).
Our approach here is based instead on chrono-projective transformations \cite{DThese,5Chrono} in the context of E-D lifts to Bargmann manifolds, which are gravitational wave spacetimes \cite{Eisenhart,Bargmann,DGH91}. Also ref.  
 \cite{Igata}  discusses similar issues.

\vskip-4mm
\begin{acknowledgments}  We are indebted to Gary Gibbons for advice and Bruno Nachtergaele for  correspondance. ME thanks the 
\emph{Denis Poisson Institute of Orl\'eans~-~Tours University} for a post-doctoral scholarship. 
 This work was partially supported by 
  National Natural Science Foundation of China (Grant No. 11575254).
P.K. was supported by the grant 2016/23/B/ST2/00727 of the National Science Centre of Poland. 
\end{acknowledgments}
\goodbreak



\begin{thebibliography}{99} 

\bibitem{LLMech}
L. Landau and E. Lifchitz,
Physique Th\'eorique, Tome I:
\textit{M\'ecanique}. $3^{e}$ \'edition. \'Editions MIR, Moscou (1969);
V. Arnold, \textit{Les m\'ethodes math\'ematiques de la 
m\'ecanique classique}, \'Editions MIR, Moscou (1976).


\bibitem{KHarmonies} 
  P.-M.~Zhang, M.~Cariglia, M.~Elbistan, G.~W.~Gibbons and P.~A.~Horvathy, 
  ``"Kepler Harmonies'' and conformal symmetries,''
 Physics Letters B 792 (2019) 324 
 [{\tt arXiv:1903.01436 [gr-qc]}].
 https://doi.org/10.1016/j.physletb.2019.03.057
  
\bibitem{Eisenhart} 
L. P. Eisenhart, ``Dynamical trajectories and geodesics", Annals Math. {\bf 30} 591-606 (1928). 
  
\bibitem{Bargmann}
C. Duval, G. Burdet, H. P. K\"{u}nzle and M. Perrin,
 ``Bargmann structures and Newton-Cartan theory",
Phys. Rev. {\bf D 31} (1985) 1841.
  
\bibitem{DGH91}
C. Duval, G.W. Gibbons, P. Horvathy,
 ``Celestial mechanics, conformal structures and gravitational waves,''
 Phys. Rev. {\bf D43} (1991) 3907. [hep-th/0512188]. 
 
\bibitem{Conf4GW}
  P.-M.~Zhang, M.~Cariglia, M.~Elbistan and P.~A.~Horvathy,
  ``Scaling and conformal symmetries for plane gravitational waves,''
  [{\tt arXiv:1905.08661 [gr-qc]}].
 
\bibitem{DThese} 
  C. Duval,
 ``Quelques proc\'edures g\'eom\'etriques en dynamique des particules,'' Doctorat
d'Etat \`es Sciences (Aix-Marseille-II), 1982 (unpublished). 
See also
  G.~Burdet, C.~Duval and M.~Perrin,
  ``Cartan Structures On Galilean Manifolds: The chrono-projective Geometry,''
  J.\ Math.\ Phys.\  {\bf 24} (1983) 1752.
  doi:10.1063/1.525927.
  
\bibitem{5Chrono}
  M.~Perrin, G.~Burdet and C.~Duval,
  ``chrono-projective Invariance of the Five-dimensional Schr\"odinger Formalism,''
  Class.\ Quant.\ Grav.\  {\bf 3} (1986) 461.
  doi:10.1088/0264-9381/3/3/020
  
\bibitem{Cariglia15}
  M.~Cariglia,
  ``Null lifts and projective dynamics,''
  Annals Phys.\  {\bf 362} (2015) 642
  doi:10.1016/j.aop.2015.09.002
  [{\tt arXiv:1506.00714 [math-ph]}].

\bibitem{JacobiLec}
C. G. J. Jacobi, ``Vorlesungen \"uber Dynamik.''
Univ. K\"onigsberg 1842-43. Herausg. A. Clebsch.
Vierte Vorlesung: Das Princip der Erhaltung der lebendigen Kraft. Zweite ausg. 
C. G. J. Jacobi's Gesammelte Werke. Supplementband. Herausg. E. Lottner. Berlin Reimer (1884).

\bibitem{Schr}
R. Jackiw, ``Introducing scaling symmetry,''
 Phys. Today, {\bf 25} (1972) 23;
U. Niederer, 
``The maximal kinematical symmetry group of the free Schr\"odinger equation,''
 Helv. Phys. Acta {\bf 45} (1972) 802 \;
 C. R. Hagen, 
 ``Scale and conformal transformations in Galilean-covariant field theory,''
 Phys. Rev. {\bf D5} (1972) 377.
 
\bibitem{Fubini}
  V.~de Alfaro, S.~Fubini and G.~Furlan,
  ``Conformal Invariance in Quantum Mechanics,''
  Nuovo Cim.\ A {\bf 34} (1976) 569.
  doi:10.1007/BF02785666
 
\bibitem{Henkel} 
 M. Henkel,
``Local Scale Invariance and Strongly Anisotropic Equilibrium Critical Systems,''
Phys Rev. Lett. {\bf 78} (1997), 1940 ;
  ``Phenomenology of local scale invariance: from
  conformal invariance to dynamical scaling,''
  Nucl.\ Phys.\  B {\bf 641} (2002) 405
  
\bibitem{DHNC}
 C.~Duval and P.~A.~Horvathy,
  ``Non-relativistic conformal symmetries and Newton-Cartan structures,''
 J.\ Phys.\ A {\bf 42} (2009) 465206
  doi:10.1088/1751-8113/42/46/465206
  [{\tt arXiv:0904.0531 [math-ph]}]
  
   
\bibitem{HallSteele}  
  G.~S.~Hall and J.~D.~Steele,
  ``Conformal vector fields in general relativity,''
  J.\ Math.\ Phys.\  {\bf 32} (1991) 1847.
  doi:10.1063/1.529249 ;
  G.~S.~Hall, ``Symmetries and Curvature Structure in General Relativity'', World Scientific (2004). 
 
\bibitem{Brinkmann}
M. W. Brinkmann,
 ``Einstein spaces which are mapped conformally on each other,''
 Math. Ann. {\bf 94} (1925)~119--145.       

\bibitem{Brdicka}
  M. Brdi\v{c}ka, 
 ``On Gravitational Waves,'' 
Proc. Roy. Irish Acad. 54A (1951) 137.

\bibitem{Torre}
C.~G.~Torre,
``Gravitational waves: Just plane symmetry,''
Gen.\ Rel.\ Grav.\  {\bf 38} (2006) 653
doi:10.1007/s10714-006-0255-8
[{\tt gr-qc/9907089}].

\bibitem{AndrPrenc}
  K.~Andrzejewski and S.~Prencel,
  ``Memory effect, conformal symmetry and gravitational plane waves,''
  Phys.\ Lett.\ B {\bf 782} (2018) 421
  doi:10.1016/j.physletb.2018.05.072
  [{\tt arXiv:1804.10979 [gr-qc]}].
 
\bibitem{Carroll4GW}
 C.~Duval, G.~W.~Gibbons, P.~A.~Horvathy and P.-M.~Zhang,
``Carroll symmetry of plane gravitational waves,''
Class. Quant. Grav. {\bf 34} (2017).
doi.org/10.1088/1361-6382/aa7f62. [{\tt arXiv:1702.08284 [gr-qc]}].

\bibitem{POLPER}
  P.~M.~Zhang, C.~Duval, G.~W.~Gibbons and P.~A.~Horvathy,
 ``Velocity Memory Effect for Polarized Gravitational Waves,''
  JCAP {\bf 1805} (2018) no.05,  030
  doi:10.1088/1475-7516/2018/05/030
  [{\tt arXiv:1802.09061 [gr-qc]}].

\bibitem{Ilderton}
  A.~Ilderton,
  ``Screw-symmetric gravitational waves: a double copy of the vortex,''
  Phys.\ Lett.\ B {\bf 782} (2018) 22
  doi:10.1016/j.physletb.2018.04.069
  [{\tt arXiv:1804.07290 [gr-qc]}].
  
  
\bibitem{Dimakis}
  N.~Dimakis, P.~A.~Terzis and T.~Christodoulakis,
  ``Integrability of geodesic motions in curved manifolds through non-local conserved charges,''
    Phys.\ Rev.\ D {\bf 99} (2019) no.10,  104061
  doi:10.1103/PhysRevD.99.104061
 [{\tt arXiv:1901.07187 [gr-qc]}].

     
\bibitem{vKampen}  
N.G. Van Kampen, 
``Transformation Groups and the Virial Theorem,'' Reports on Math. Physics {\bf 3}, 235 (1972).
  
\bibitem{Nachtergaele}
  B. Nachtergaele and A. Verbeure, 
``Groups of canonical transformations and the virial-Noether theorem,'' Journal of Geometry and Physics, 3, 315-325 (1986)
 
\bibitem{Keplar}
  N.~Ogawa,
  ``A Note on the scale symmetry and Noether current,''
  hep-th/9807086.
  
\bibitem{Igata}
  T.~Igata,
  ``Scale invariance and constants of motion,''
  PTEP {\bf 2018} (2018) no.6,  063E01
  doi:10.1093/ptep/pty060
  [arXiv:1804.03369 [hep-th]].
 
\end{thebibliography}
\end{document}